\documentclass[11pt,a4paper]{article}                           

\usepackage{amsmath}
\usepackage{amssymb}
\usepackage[usenames]{color} 
\usepackage{multirow}
\usepackage{graphicx}
\usepackage{epstopdf}
\usepackage{array}
\usepackage{hhline}
\usepackage{cite}
\usepackage{microtype,colortbl}
\usepackage[bf]{caption}
\usepackage{color}
\usepackage[usenames,dvipsnames]{xcolor}
\usepackage{pstool}
\usepackage{tikz}
\usepackage{ytableau}
\usepackage{latexsym,stmaryrd}
\usepackage[bookmarks=false]{hyperref}
 \usepackage{lscape}
 \usepackage{subfig}

\usepackage{enumitem}

\usepackage{ifpdf}
\def\hybrid{\topmargin -20pt    \oddsidemargin 0pt
        \headheight 0pt \headsep 0pt
        \textwidth 6.25in       
        \textheight 9 in       
        \marginparwidth .875in
        \parskip 5pt plus 1pt 
          \jot = 1.5ex
   }
\hybrid
\numberwithin{equation}{section}
\numberwithin{table}{section}

\newcommand{\beq}{\begin{equation}}  \newcommand{\eeq}{\end{equation}}
\newcommand{\bal}{\begin{aligned}}   \newcommand{\eal}{\end{aligned}}
\newcommand{\bea}{\begin{eqnarray}}  \newcommand{\eea}{\end{eqnarray}}

\newcommand{\bmat}{\left(\begin{array}}
\newcommand{\emat}{\end{array}\right)}

\newcommand{\nn}{\nonumber}

\newcommand{\cK}{\mathcal{K}}
\newcommand{\cN}{\mathcal{N}}

\newcommand{\cI}{\mathcal{I}}

\newcommand{\cV}{\mathcal{V}}

\renewcommand{\Im}{\mathrm{Im}\,}

\usepackage{cancel}

\newcommand{\be}{\begin{equation}}
\newcommand{\ee}{\end{equation}}

\definecolor{Gray}{gray}{0.95}

\begin{document}
\baselineskip=14pt
\parskip 5pt plus 1pt 

\vspace*{2cm}
\begin{center}
{\Large \bfseries Engineering Small Flux Superpotentials \\[.3cm]
and Mass Hierarchies}\\[.3cm]

\vspace{1cm}
{\bf Brice Bastian}\footnote{b.bastian@uu.nl},
{\bf Thomas W.~Grimm}\footnote{t.w.grimm@uu.nl},
{\bf Damian van de Heisteeg}\footnote{d.t.e.vandeheisteeg@uu.nl},

{\small
\vspace*{.5cm}
Institute for Theoretical Physics, Utrecht University\\ Princetonplein 5, 3584 CC Utrecht, The Netherlands\\[3mm]
}
\end{center}
\vspace{1cm}
\begin{abstract} \noindent
We study the stabilization of complex structure moduli in Type IIB flux compactifications by using recent general results 
about the form of the superpotential and K\"ahler potential near the boundaries of the moduli 
space. In this process we show how vacua with an exponentially small vacuum superpotential can be realized systematically and 
understood conceptually within asymptotic Hodge theory. 
We distinguish two types of vacua realizing such superpotentials 
that differ by the mass scales of the stabilized moduli. 
Masses polynomially depending on the moduli arise if the superpotential contains exponential corrections whose existence is required to ensure the non-degeneracy of the moduli space metric.
We use the fact that such essential corrections are controlled by asymptotic Hodge theory and have 
 recently been constructed for all one- and two-moduli asymptotic regimes. 
These insights allow us to obtain new vacua near boundaries in complex structure moduli space that include Seiberg-Witten points. In these examples we find that the scale of the vacuum superpotential can be bounded from below through the exponential of the negative D3-brane tadpole.
\end{abstract}

\newpage

\setcounter{footnote}{0}

\tableofcontents
\newpage
\section{Introduction}

Flux compactifications of Type IIB string theory provide some of the best understood scenarios that 
can yield minimally supersymmetric or non-supersymmetric four-dimensional effective actions with partially or fully
stabilized moduli \cite{Grana:2005jc,Douglas:2006es}. These scenarios use the fact that background R-R and NS-NS three-form fluxes introduce 
a non-trivial superpotential that depends on the complex structure deformations of the internal compactification space 
and the axio-dilaton. Remarkably, the backreaction of these fluxes on the geometry yields at leading order only a 
warp factor and the requirement to consider a conformal Calabi-Yau background combined with an orientifold projection. 
The effective theories can thus be studied by appropriately extending the techniques used to describe the geometry of Calabi-Yau threefolds and their moduli spaces.
This has lead to the suggestion of concrete moduli stabilization 
scenarios in \cite{Giddings:2001yu,Kachru:2003aw,Balasubramanian:2005zx}. The search for explicit realizations of these scenarios still tests our 
understanding of string compactifications and pushes the boundaries of what is possible or impossible within string theory. 

In this work we will investigate the stabilization of complex structure moduli after introducing background fluxes. 
Following the well-known dimensional reduction on a Calabi-Yau orientifold \cite{Grana:2005jc,Douglas:2006es,Giddings:2001yu,Grimm:2004uq}, we recall that the effective theory is an $\cN=1$ supergravity theory. The for us relevant part of the $\cN=1$ superpotential and K\"ahler potential can be evaluated by determining how the (up to rescaling) unique $(3,0)$-form $\Omega$ on the Calabi-Yau threefold $Y_3$ varies with a change of the complex structure. This dependence can be captured by so-called period integrals, or periods for short, which are obtained by integrating $\Omega$ over a basis of three-cycles of $Y_3$.
There are a number of techniques available to derive these period integrals. 
While most of them require the explicit construction of the Calabi-Yau manifold, it was recently shown in \cite{Bastian:2021eom}
that general models for the asymptotic periods can be constructed by asymptotic Hodge theory. In a nutshell, this 
mathematical formalism uses the existence of monodromy symmetry, the requirement of positivity and orthogonality, and a notion of completeness. It then allows for the construction of the compatible periods with these properties in an algorithmic way 
without resorting to any specific geometric examples. With these powerful techniques at hand we aim to 
understand how certain properties of the effective theory, such as the vacuum expectation value 
of the superpotential and mass hierarchies of the stabilized moduli can be engineered algorithmically. 

One of the most prominent moduli stabilization scenarios, the KKLT scenario \cite{Kachru:2003aw}, requires to find vacua in 
complex structure moduli space, such that the vacuum superpotential is taking a very small, non-zero value.  
It was recently suggested in \cite{Demirtas:2019sip} and further explored in \cite{Demirtas:2020ffz,Blumenhagen:2020ire,Honma:2021klo,Demirtas:2021nlu, Demirtas:2021ote,Broeckel:2021uty} that, in fact, exponentially small vacuum 
superpotentials can be found near certain boundaries in moduli space.\footnote{See \cite{Denef:2004ze, Eguchi:2005eh} for statistical arguments about the abundance of vacua in these regions.} It was argued that the construction proceeds 
by introducing fluxes that preserve a continuous version of the monodromy symmetry at some leading order with a vanishing 
superpotential, while then including instanton corrections generates an exponentially small superpotential. We will 
explain how this construction is understood in asymptotic Hodge theory and how it can be generalized to other boundaries that 
have not been consider in the literature before. This will be done both in a hands-on way by considering explicit examples as well as by explaining how the abstract mathematical methods are useful. 
We will then see that our powerful techniques also allow us to control the moduli masses and 
identify alternative scenarios to the ones of \cite{Demirtas:2019sip,Demirtas:2020ffz}. In our constructions the scaling 
of the masses is polynomial, rather than exponential, in the vacuum expectation values of the moduli. 
In general, the Hodge theory methods enable us to highlight conceptual 
features of these moduli stabilization scenarios and show how to implement them as a building block on general moduli spaces. 
In the mathematical literature, the vacua with 
exponentially small superpotential are a special case of the so-called
extended locus of Hodge classes \cite{schnell2014extended}. 

To highlight some key aspects of our construction let us first point out that asymptotic Hodge theory gives a 
general way of modeling the asymptotic periods of $\Omega$. Firstly, there is the famous result \cite{Schmid}
that ensures that the leading polynomial part of the periods is captured by the monodromy symmetry and some 
data associated to the asymptotic regime. For example, in the large complex structure regime of a 
one-dimensional moduli space the asymptotic periods can be written as $\Pi_{\rm pol} = e^{tN} a_0 
$, where $t$ is the complex structure modulus with $t = i\infty$ being the large complex structure point, $N$ is the logarithm 
of the monodromy matrix and nilpotent $N^4=0$, and $a_0$ is a vector associated to the limit. This form naturally generalizes to 
all limits in complex structure moduli space in one or more directions. Secondly, it was shown in \cite{Bastian:2021eom} that away from the large complex structure regime 
exponential corrections to $\Pi_{\rm pol} $ are needed both for mathematical consistency as well as for the physical effective theory to be well-defined.  
We will thus split the asymptotic periods as $\Pi_{\rm pol} + \Pi_{\rm ess}$ and refer to the corrections $\Pi_{\rm ess}$ as \textit{essential instantons}. This name can be motivated by 
mirror symmetry, where at the large complex structure point exponential corrections actually map to instanton corrections in the mirror theory.  
The necessity of these corrections can be seen, for example, when the metric derived solely from $\Pi_{\rm pol}$  degenerates. Remarkably, the 
asymptotic Hodge theory techniques introduced  in \cite{Schmid,CKS,CKAsterisque} can then be used to determine the general form of the needed  exponentially suppressed corrections $\Pi_{\rm ess}$
to exactly avoid such inconsistencies.

With these insights about the structure of the periods we  search for flux vacua that admit exponentially small vacuum superpotentials. 
Using the classification of asymptotic regimes \cite{Kerr2017,Grimm:2018cpv,Grimm:2019bey} we are able to identify a specific class of regions that 
generally admits essential instanton corrections that can set the scale of the vacuum superpotential. 
We find that a large class of such regions arise near boundaries containing Seiberg-Witten points. These special 
loci occur in the moduli space of Calabi-Yau geometries that are used to embed Seiberg-Witten theory into type IIB string theory \cite{Kachru:1995fv,Eguchi:2007iw}. 
Among the set of essential instantons in the flux superpotential we then identify a subset that enters the 
scalar potential already at polynomial order. These are 
precisely the instanton corrections that are required to ensure a non-degenerate moduli metric and induce  the polynomial behavior of the moduli masses in the vacuum. We refer to these corrections as \textit{metric-essential instantons} in this paper. 
Furthermore, we then find that in these vacua the canonically 
normalized masses of the axio-dilaton and complex structure moduli can be parametrically heavier 
compared to the mass of the K\"ahler moduli.

 In order to illustrate our method, we then specialize our analysis to one and two moduli settings. General models for the asymptotic periods of these low-dimensional settings have recently been constructed in \cite{Bastian:2021eom} and we will use them here to perform explicit computations. It turns out that these concrete realizations lead to additional intriguing observations about the hierarchies among moduli masses and the minimal scale of the vacuum superpotential. While we are not providing any general proofs on these findings, they are suggestive of a deeper structure and provide additional motivation for an explicit study of higher dimensional examples. Finally, we should mention that the period models used in this work are given in a real rather than a rational symplectic basis. Our vacua do not require any fine-tuning of the flux quanta in contrast to e.g.~a racetrack potential, so we do not expect any technical issues to arise here. Nevertheless it would be interesting to work out the quantization of the fluxes in the future.

The paper is organized as follows. In section \ref{sec:IIBflux} we review some basic aspects of Type IIB flux compactifications, and outline our strategy for constructing vacua with exponentially small superpotentials. Particularly, we review how essential instantons can play a crucial role in e.g.~the K\"ahler metric, and explain how to carefully deal with these corrections in the context of moduli stabilization. In section \ref{sec:examples}, to illustrate this story further, we then consider two simple examples, where we show how to stabilize the moduli explicitly such that we engineer an exponentially small vacuum superpotential. In section \ref{sec:conclusions} we present our conclusions and give promising directions for future research.

\section{Type IIB flux compactifications and small superpotentials}\label{sec:IIBflux}
In this section we stepwise explain the general construction of flux vacua with small vacuum superpotentials. We start by briefly reviewing the basics of four-dimensional supergravities arising from type IIB orientifold compactifications. Then we recall some of the main findings of \cite{Bastian:2021eom}, namely that exponential corrections to the asymptotic periods are required away from the large complex structure regime and how these can contribute at polynomial level to the scalar potential. Using this knowledge, we outline a method to construct a set of flux vacua with exponentially small vacuum superpotential near a given asymptotic regime. The reader interested only in concrete examples can safely skip to section 
\ref{sec:examples}.

\subsection{Review: flux vacua in Type IIB orientifold compactifications}
Let us begin with a brief review of some generalities on Type IIB flux compactifications, see e.g.~\cite{Grana:2005jc,Douglas:2006es} for a more complete overview. We consider compactifications of Type IIB string theory on a Calabi-Yau threefold $Y_3$, which is subjected to an orientifold involution. We focus on the complex structure and axio-dilaton sectors of the four-dimensional $\mathcal{N}=1$ supergravity theories arising from these compactifications, where the orientifold projection might freeze out some of the $h^{2,1}(Y_3)$ complex structure moduli. For the purposes of our work we can ignore the K\"ahler moduli, so let us write the K\"ahler potential as
\begin{equation}
K = -\log[i(\bar{\tau} - \tau)] - \log\Big[i \int_{Y_3}  \bar \Omega(\bar{t}^i) \wedge \Omega(t^i) \, \Big]\, , \label{eq:KP}
\end{equation}
where $\Omega$ is the (up to rescaling) unique $(3,0)$-form of $Y_3$, and we denoted the axio-dilaton and complex structure moduli respectively by
\begin{equation}\label{eq:coordinates}
\tau=c+i s\, , \qquad t^i = x^i + i y^i\, , \qquad i = 1,\ldots , h^{2,1}_-(Y_3)\, ,
\end{equation}
where $h^{2,1}_-(Y_3)$ is the number of complex structure moduli that survived the orientifold projection. As will become more clear later we refer throughout this work to the real part of the moduli $c, x^i$ as the physical axions, while we identify their imaginary counterparts $s,y^i$ as the saxions. 

In order to make the story more explicit, it proves to be convenient to write out $\Omega$ in terms of its periods. For this purpose let us introduce a basis $\gamma_\mathcal{I} \in H^3_-(Y_3, \mathbb{R})$ for the orientifold-odd threefold cohomology. We can expand $\Omega$ in this basis in its periods $\Pi^\cI$ as
\begin{equation}\label{eq:periods}
\Omega = \Pi^\mathcal{I} \gamma_\mathcal{I} \, , \qquad \mathcal{I} = 0\, ,\ \ldots\, , \ 2h^{2,1}_-(Y_3)+2\, .
\end{equation}
This allows us to write the K\"ahler potential for the complex structure moduli in terms of the periods as
\begin{equation}\label{eq:kahler}
K^{\rm cs} = - \log \big[  i  \bar \Pi^T \eta\, \Pi  \big] =  - \log \big[  i \langle \bar \Pi  ,  \Pi \rangle  \big] \, , \qquad \\
\end{equation}
where we introduced the symplectic pairing 
\begin{equation}
\langle u , \,  v \rangle  =  \int_{Y_3} u \wedge v  = u^T \eta \, v \, , \qquad  \eta = \begin{pmatrix} 0 & \mathbb{I} \\
- \mathbb{I} & 0 
\end{pmatrix}\, ,
\end{equation}
with $\mathbb{I}$ the $(h^{2,1}_-+1)$-dimensional identity matrix. The standard K\"ahler metric is conveniently expressed in terms of the K\"ahler potential \eqref{eq:KP} as 
\begin{align}
K_{I \bar{J}}=\partial_{I} \bar{\partial}_{\bar{J}} K\, ,
\end{align}
where the indices $I,J$ run over the axio-dilaton $\tau$ and the complex structure moduli $t^i$. Turning on R-R and NS-NS fluxes $F_3$ and $H_3$, we then induce a superpotential in the four-dimensional $\mathcal{N}=1$ effective theory. This flux superpotential can be written in terms of $\Omega$ as \cite{Gukov:1999ya}
\begin{equation}\label{eq:superpotential}
W = \int_{Y_3} G_3 \wedge \Omega = \langle G_3 \, , \Pi \rangle \, , \qquad G_3=F_3- \tau H_3\, .
\end{equation} 
By means of standard $\mathcal{N}=1$ identities one can then write the scalar potential in terms of this superpotential. Using the no-scale property of the K\"ahler moduli, see for instance \cite{Grimm:2004uq} for the details, we can write this scalar potential out as
\begin{equation}\label{eq:potential}
V = \frac{1}{\cV^2 }e^K  K^{I \bar{J}} D_I W D_{\bar{J}} \bar{W} = \frac{1}{4 \cV^2\, s}  \big( \langle \bar{G}_3\, , \ \ast G_3 \rangle - i \langle  \bar{G}_3\ , \, G_3  \rangle \big)\, ,
\end{equation} 
where $K^{I \bar{J}}$ denotes the inverse of the K\"ahler metric, and  $D_I W = \partial_I W+K_I W$. We denoted the volume factor depending on the K\"ahler moduli of $Y_3$ by $\cV$, which will not be relevant to our work. 

One can now study the minima of this flux potential in two equivalent ways \cite{Giddings:2001yu,Grimm:2004uq}. The standard $\cN=1$ supergravity approach is to solve for vanishing F-terms, yielding as constraints
\begin{equation}\label{eq:Fterms}
D_I W = \partial_I W + K_I W =  0\, .
\end{equation}
Alternatively one considers the imaginary self-duality condition for the three-form flux $G_3$, which reads
\begin{equation}\label{eq:selfduality}
\ast G_3 = i G_3\, .
\end{equation}
From both approaches one sees that the scalar potential \eqref{eq:potential} vanishes at the minimum, giving rise to a Minkowski-type vacuum. In our construction of flux vacua in section \ref{ssec:strategy} and \ref{sec:examples} we choose to work mostly with the vanishing F-term conditions \eqref{eq:Fterms}. However, as becomes clear from our discussion of exponential corrections in section \ref{ssec:asymptotics}, it is instructive to consider the self-duality condition \eqref{eq:selfduality} as well as a complementary perspective.

As a final remark, we recall that the three-form fluxes $F_3, H_3$ contribute to the D3-brane tadpole. This contribution due to the fluxes is given by
\begin{equation}\label{eq:tadpole}
Q_{\rm flux} = \frac{1}{2} \langle F_3, \,  H_3 \rangle\, .
\end{equation}
The cancellation condition for the D3-brane tadpole then puts an upper bound on the allowed flux $Q_{\rm flux}$.

\subsection{Near-boundary asymptotics and exponential corrections}
\label{ssec:asymptotics}
In this section we discuss the characteristic features of these four-dimensional effective theories near boundaries in complex structure moduli space. To be more precise, we consider the asymptotic behavior of the K\"ahler potential \eqref{eq:kahler}, flux superpotential \eqref{eq:superpotential} and scalar potential \eqref{eq:potential} in these regimes. We in particular explain why near certain boundaries some exponentially small corrections cannot be dropped in the former two (\eqref{eq:kahler} and \eqref{eq:superpotential}), while a polynomial approximation for the scalar potential \eqref{eq:potential} is sufficient. This discussion of exponential corrections is based on our recent observations in \cite{Bastian:2021eom}, which we refer to for a more detailed treatment of these subtleties. In the following we formulate these main findings in a general setting, and in sections \ref{ssec:II0} and \ref{ssec:II1} we discuss them in practice for two explicit examples. For closely related applications of these asymptotic Hodge theory techniques in the context of string compactifications we refer the reader to \cite{Grimm:2018ohb,Grimm:2018cpv,Corvilain:2018lgw,Font:2019cxq,Grimm:2019wtx,Grimm:2019ixq,Gendler:2020dfp,Lanza:2020qmt,Grimm:2020cda,Bastian:2020egp,Calderon-Infante:2020dhm,Grimm:2020ouv,Grimm:2021ikg,Lanza:2021qsu,Castellano:2021yye,Palti:2021ubp}.

Let us begin by establishing how we parametrize the complex structure moduli space and its boundaries. In this work we will employ two coordinate patches to describe boundaries. The first patch defines a codimension $n$ boundary as the normal intersection of $n$ loci of the form $z^i=0$, such that locally the boundary is given by $z^1  = \ldots = z^n = 0$. In these coordinates circling the boundaries corresponds to multiplying by a phase as $z^i \to e^{2\pi i} z^i$. Often one finds that the K\"ahler potential does not depend on the phase of the $z^i$ near these boundaries, so it proves to be useful to introduce another set of coordinates to exploit this feature. We already mentioned these coordinates briefly in \eqref{eq:coordinates}, and the precise relation between these patches is given by
\begin{equation}
z^i = e^{2\pi i t^i} = e^{2\pi i( x^i + i y^i)}\, .
\end{equation}
From a physical perspective one can interpret the $x^i$ that parametrize the phase of $z^i$ as \textit{axions}, where circling the boundary now corresponds to a discrete shift symmetry $x^i \to x^i+1$. Their counterparts $y^i$ are then referred to as \textit{saxions}, which are the fields we make large as $y^i \to \infty$ in order to approach the near-boundary region.

The asymptotic behavior of the K\"ahler potential and flux superpotential is then best understood by expanding the periods \eqref{eq:periods} around a boundary in complex structure moduli space.  The nilpotent orbit theorem \cite{Schmid} tells us that this period vector admits an expansion
\begin{equation}\label{eq:periodsexpansion}
 \Im t^i = y^i \gg 1\, : \qquad \Pi(t^i)  = e^{t^i N_i} \big( a_0 + \sum_{r_i } e^{2\pi i r_i t^i} a_{r_1 \cdots r_n}  \big)\, ,
\end{equation}
where the $N_i$ denote a set of mutually commuting matrices whose form depends on the monodromy behavior when circling the boundary and the terms $a_{r_1\ldots r_n}$ are independent of the coordinates $t^1, \ldots, t^n$ taken close to the boundary. The first term $a_0$ will give rise to polynomial terms in $y^i$ in the K\"ahler potential \eqref{eq:kahler} and superpotential \eqref{eq:superpotential}, while the other $a_{r_1\ldots r_n}$  lead to exponentially suppressed terms. Borrowing the nomenclature familiar from large complex structure and applying it to any boundary in complex structure moduli space, we loosely refer to $a_0$ as the perturbative term, while the other exponentially suppressed terms resemble the instanton expansion.

A remarkable feature of asymptotic Hodge Theory is that is provides an exhaustive classification of all possible boundaries that can occur in complex structure moduli space. In this work, we only need the broad lines of this classification and refer the reader to \cite{Kerr2017,Grimm:2018cpv,Grimm:2019bey} for more details. In the case of Calabi-Yau threefolds, there are four main types of boundaries which are conventionally labeled by the Greek numerals $\rm I , II , III$ and $\rm IV$. Each of these has a finite number of subtypes that we usually label by a subscript. However, this level of detail is not required in what follows, so we suppress this additional label and work only with the main types. These can be assigned to each co-dimension $n$ boundary by studying the associated monodromy properties of the period vector in more detail. For our study of flux vacua we can limit ourselves simply to the types associated with the co-dimension $h^{2,1}$ boundary, i.e. points where all the moduli have been sent to the limit. Given a co-dimension $h^{2,1}$ locus, we can characterize the different types by the action of the operator $N=N_1 + \dots + N_{h^{2,1}}$ on $a_0$ as follows
\begin{align}
{\rm Type \, I}:& \quad N a_0 =0\,, \nonumber \\
{\rm Type \, II}:& \quad N^2 a_0 =0\,, \nonumber \\
{\rm Type \, III}:& \quad N^3 a_0 =0\,, \nonumber \\
{\rm Type \, IV}:& \quad N^4 a_0 =0  \,,
\end{align}
where it is understood that $N^{\bullet-1}a_0 \neq 0$ for a given type.
Let us make some general remarks to give the reader a better feeling for the different cases. The type $\rm I$ are the class of finite distance boundaries, while the others are an infinite distance away when measuring in the moduli space metric. The most famous example of a type $\rm I$ boundary is the conifold point \cite{Strominger_1995}. The type $\rm II$ boundaries signal the presence of a K3 fibration in the mirror geometry \cite{doran2016mirror} and are thus relevant for the geometric engineering of Seiberg-Witten theories \cite{Kachru:1995fv,Eguchi:2007iw}. Type $\rm III$ boundaries signal the presence of an elliptic fibration at large complex structure, but cannot occur in other regions in moduli space for $h^{2,1}<3$. Consequently, they remain less well studied (at least in the context of flux vacua) due to the increased complexity of even the simplest occurrence away from LCS. The remaining type $\rm IV$ boundaries can be further separated into two groups. On one side there are the well known large complex structure (LCS) points, which are the most studied due to their relatively simple form of their periods and the relevance for mirror symmetry. On the other side there is the group of the so called coni-LCS boundaries, which have recently been studied by \cite{Blumenhagen:2020ire,Demirtas:2020ffz} in the context of flux compactification.

One of the main findings in \cite{Bastian:2021eom} is that asymptotic Hodge theory requires the presence of instanton terms $a_{r_1\ldots r_n}$ when considering boundaries that are \textit{not} of large complex structure type. This argument uses the fact that the derivatives of the period vector of $\Omega$ should span the complete three-form cohomology $H^3(Y_3, \mathbb{C})$. At large complex structure it suffices to consider the descendants $a_0, N_i a_0, N_i N_j a_0, N_i N_j N_k a_0$ of the perturbative term, but for other boundaries it was shown that this set of vectors does not span a $2(h^{2,1}+1)$-dimensional space. To reconcile this issue we have to include a finite number of terms $a_{r_1 \cdots r_n}$ in the expansion \eqref{eq:periodsexpansion}, which we therefore refer to as \textit{essential instantons}. 

In order to make the necessity of these essential instantons clear from a physical perspective, it is instructive to study the K\"ahler metric. In this context one needs the K\"ahler metric to be non-degenerate in order to have properly defined kinetic terms in the four-dimensional effective action. By inserting the expansion \eqref{eq:periodsexpansion} back into the K\"ahler potential \eqref{eq:kahler}, we can split the contributions from the periods into two parts
\beq \label{eq:kahlerexpansion}
K^{\rm cs} = - \log[\cK_{\rm pol} + \cK_{\rm inst} ] \, , 
\eeq
with
\beq
\cK_{\rm pol} =i \langle \bar a_0 , e^{2i y^i N_i} a_0 \rangle  \, , \qquad \cK_{\rm inst} =  \sum_{r_i, s_i \geq 0}  e^{-2\pi [  y^i (s_i +r_i)+i x^i( s_i - r_i)]}i \langle \bar  a_{r_1 \ldots r_n}  , e^{2i y^i N_i} a_{s_1 \ldots s_n}  \rangle  \, , \nn 
\eeq
where the sum runs over integers $r_i,s_i \geq 0$ with either $r_i \neq 0$ or $s_i \neq 0$. This partition separates the purely polynomial terms in the K\"ahler potential from the exponentially suppressed terms. The crucial observation is now that the perturbative metric resulting from the polynomial part $\cK_{\rm pol} $ can be degenerate, i.e.~it has a vanishing eigenvalue. For example at the conifold point one finds that $\cK_{\rm pol} $ is simply a constant, and therefore the K\"ahler potential depends on the complex structure moduli only through the exponentially suppressed part $\cK_{\rm inst} $. Furthermore, this degeneracy of the perturbative K\"ahler metric is not restricted to such finite distance boundaries, but also arises at certain infinite distance boundaries, e.g.~the class of Seiberg-Witten points we consider in section \ref{ssec:II1}. For these infinite distance class boundaries one finds that $\cK_{\rm pol} $ is at most linear in the saxions, and one can straightforwardly show that this linear behavior results in a degenerate K\"ahler metric for two or more moduli, as demonstrated below with \eqref{eq:example_metric}.

One important sidenote we should make is that a non-degenerate K\"ahler metric only requires the first derivatives of $\Omega$ to span a $h^{2,1}$-dimensional space for $H^{2,1}(Y_3, \mathbb{C})$, but higher-order derivatives relevant for $H^{1,2}(Y_3, \mathbb{C})$ and $H^{0,3}(Y_3, \mathbb{C})$ are not needed. Therefore the mathematical notion of completeness that requires the derivatives of $\Omega$ to span the three-form cohomology $H^3(Y_3, \mathbb{C})$ leads to a larger set of essential instantons than just those required by the K\"ahler metric. In other words, the K\"ahler metric generically only explains the presence of a subset of the essential instantons, which we will refer to as \textit{metric-essential}. 

To illustrate this discussion on metric-essential instantons, let us consider a simple two-moduli example. We borrow the K\"ahler potential \eqref{eq:kpII1} that describes the asymptotic regime near a type $\mathrm{II}$ point, which will be studied in more detail later. Its polynomial part reads
\begin{equation}\label{eq:example_potential}
K^{\rm cs}_{\rm pol} = - \log \cK_{\rm pol} = -\log[y_1+n_2 y_2]\, .
\end{equation}
By making a holomorphic change of variables $(t_1',t_2')= (t_1+n_2 t_2, n_2 t_1- t_2)$ one easily checks that the dependence on $y_2'$ drops out. We can equivalently see this degeneracy by explicitly computing the K\"ahler metric
\begin{equation}\label{eq:example_metric}
K_{i\bar{j}} = \frac{1}{(y_1+n_2 y_2)^2} \begin{pmatrix} 1 & n_2 \\
n_2 & (n_2)^2
\end{pmatrix}\, .
\end{equation}
Its determinant vanishes, so taking just \eqref{eq:example_potential} as K\"ahler potential leads to ill-defined kinetic terms for the complex structure moduli in this asymptotic regime. To be more precise, the eigenvector $(1,n_2)$ has a polynomial eigenvalue, while $(n_2, -1)$ has a vanishing eigenvalue. By requiring the presence of exponential corrections to \eqref{eq:example_potential} we can cure this degeneracy, resulting in an exponentially small eigenvalue for $(n_2, -1)$ instead. These corrections are precisely the metric-essential instantons which are included in the K\"ahler potential by $\cK_{\rm inst}$ in \eqref{eq:kpII1}. 

We next turn to the flux superpotential \eqref{eq:superpotential}. Similar to the K\"ahler potential we can use the expansion \eqref{eq:periodsexpansion} for the periods, and separate the terms in the superpotential into two parts
\begin{equation}\label{eq:Wexpansion}
\begin{aligned}
W &= W_{\rm pol} + W_{\rm inst} \, , \\
W_{\rm pol} &= \langle G_3\, , \ e^{t^i N_i} a_0 \rangle\, , \qquad  &W_{\rm inst} &=\sum_{r_i } e^{2\pi i r_i t^i}  \langle G_3\, ,  \  e^{t^i N_i} a_{r_1 \cdots r_n}  \rangle  \, ,
\end{aligned}
\end{equation}
where the sum over $r_i$ runs over at least one $r_i \neq 0$. Similar to the expansion of the K\"ahler potential \eqref{eq:kahlerexpansion}, essential instantons in the periods can play an important role for the superpotential, meaning one cannot drop all terms in $W_{\rm inst}$ near the boundary. For instance, one can find that some flux quanta only enter the superpotential through $W_{\rm inst}$, as is the case for the class of Seiberg-Witten boundaries considered in section \ref{ssec:II1}. 

Finally, let us get to the asymptotic behavior of the scalar potential \eqref{eq:potential}. This is most easily understood by using the formulation in terms of the Hodge star operator, since its near-boundary behavior is relatively simple compared to the F-terms of the above superpotential. By performing a near-boundary expansion for the Hodge star one finds a leading polynomial piece and subleading exponential corrections, similar to our previous approximations. However, a crucial observation about this asymptotic behavior is that the polynomial part of the Hodge star operator is already non-degenerate. In contrast to K\"ahler potential \eqref{eq:kahlerexpansion}, one can therefore safely drop exponential corrections to the Hodge star without losing essential information. Furthermore, the non-degeneracy of the polynomial Hodge star means that all fluxes will enter in the scalar potential already at polynomial order.

This last observation appears to be in contradiction with our study of the superpotential \eqref{eq:Wexpansion} at first sight, since from this perspective we found some fluxes that only appears at exponential order. In order to resolve this seeming conundrum it is instructive to recall that the scalar potential \eqref{eq:potential} is computed by contracting the inverse K\"ahler metric $K^{I \bar J}$ with covariant derivatives of the superpotential $D_I W$ and its conjugate. Now if the polynomial K\"ahler metric is degenerate, then essential corrections in $\cK_{\rm inst}$ resolve this singular behavior, yielding exponentially small eigenvalues. Conversely, one finds that the inverse K\"ahler metric must have some exponentially large eigenvalues. For the scalar potential this means that the scaling of essential exponential corrections in $D_I W_{\rm inst}$ is cancelled off by these eigenvalues of the inverse K\"ahler metric $K^{I \bar J}$, such that all fluxes appear at polynomial order. In other words, these exponential terms in the superpotential produce polynomial terms in the scalar potential, so these essential corrections in $W_{\rm inst}$ \textit{cannot} be dropped at the leading perturbative level.\footnote{Similar to the sidenote on the K\"ahler metric, only metric-essential corrections needed to span $H^{2,1}(Y_3, \mathbb{C})$ with derivatives of $\Omega$ are relevant for the superpotential. Other essential instantons do produce exponentially suppressed terms in the scalar potential, and can therefore be ignored at the perturbative level.}

\subsection{Engineering vacua with small $W_0$}
\label{ssec:strategy}
We now lay out our strategy for engineering vacua with a small flux superpotential. This construction relies on first searching for solutions to the F-term equations coming from the polynomial periods $\Pi_{\rm pol}$ and supplementing them accordingly with constraints involving essential instantons in order to stabilize all moduli. As a result we will have $W_{\rm pol}=0$ and consequently an exponentially small vacuum superpotential $W_0$. This procedure is then explicitly applied to the one- and two-moduli type $\rm II$ models in sections \ref{ssec:II0} and \ref{ssec:II1} respectively.

Let us begin by writing down the relevant extremization conditions at polynomial level for our flux vacua. As we want to engineer vacua with an exponentially small superpotential, we require the polynomial part in the expansion of the superpotential \eqref{eq:Wexpansion} to vanish
\begin{equation}
W_{\rm pol} \big|_* = 0\, ,
\end{equation}
where we used a star to denote the evaluation of moduli at their vevs. At the polynomial level of the superpotential the vanishing F-term constraints \eqref{eq:Fterms} can then be written as
\begin{equation}\label{eq:polextr}
\begin{aligned}
  \langle F_3 - \tau_* H_3 , \,  \,  e^{t^i_* N_i} N_a a_0 \rangle = 0\, , \qquad \langle H_3 , \, e^{t^i_* N_i} a_0 \rangle = 0\, ,  \qquad \langle F_3 , \, e^{t^i_* N_i} a_0 \rangle = 0\, .\\
\end{aligned}
\end{equation}
The first set of equations follows from $\partial_a W_{\rm pol} \big|_* = 0$, while the latter two follow from $\partial_\tau W_{\rm pol}  \big|_*= 0$. Note that we replaced the covariant derivative by a partial derivative for all constraints since $W_{\rm pol}\big|_*  = 0$ for the flux vacua we are interested in.

It is important to stress that these polynomial level conditions \eqref{eq:polextr} do not suffice to obtain the vacua of 
the scalar potential. The vectors $N_a a_0$ can be parallel or even vanishing resulting in a insufficient set of F-term conditions at the polynomial level. Another way to observe this is by looking at the scalar potential $V_{\rm pol}$ obtained from the complete $V$ defined in \eqref{eq:potential} 
by dropping all exponentially suppressed corrections in the final expression. We now look for a combination $\phi$ of moduli that is not constrained by 
the condition  \eqref{eq:polextr}. If such an unfixed direction $\phi$ exists, two possibilities can occur:
\begin{itemize}
\item[(1)] The direction $\phi$ is a flat direction of the polynomial scalar potential $V_{\rm pol}$. This implies that $\phi$ is massless at polynomial order, but might obtain an exponentially small mass upon including either essential or non-essential instanton corrections. 
\item[(2)]  The direction $\phi$  is \textit{not} a flat direction of the polynomial scalar potential $V_{\rm pol}$, but rather has a mass term already at polynomial order. This implies that their mass term must arise from \textit{metric-essential instantons} correcting the periods. 
\end{itemize}
Let us comment on these two cases in turn. First, note that for case (1) the field $t_*$ not appearing in $V_{\rm pol}$ might still be stabilized after including instanton corrections. If non-essential instantons are used in order to ensure the stabilization one needs 
to check if such corrections are actually present for a given Calabi-Yau geometry and implement an appropriate stabilization scheme, such as a racetrack potential. To make this concrete in explicit examples was one of the successes of \cite{Demirtas:2019sip,Blumenhagen:2020ire,Demirtas:2020ffz}. In particular, it was shown that this can be done at large complex structure point in \cite{Demirtas:2019sip}. While there are no essential instantons in this asymptotic limit, it is well-known that generically instanton corrections arise and contribute to the superpotential. Second, we note that realization of case (2) is primarily dependent only on the type of asymptotic regime, since we have a systematic classification in which asymptotic regime certain essential instanton corrections have to be present. One has thus a stabilization mechanism that is independent of the actual Calabi-Yau geometry that realizes this limit. In the following we will focus on case (2) and explain in detail how such situations can be engineered. 

To gain a better understanding of what happens in case (2), let us note that such vacua exactly arise if 
the polynomial part of the K\"ahler potential $\cK_{\rm pol}$ yields a degenerate K\"ahler metric, i.e.~when the full K\"ahler metric has exponential eigenvalues. To see this we 
recall from the discussion below \eqref{eq:Wexpansion} that such a K\"ahler metric allows exponentially suppressed contributions in the F-terms to enter the scalar potential \eqref{eq:potential} at polynomial order. Hence, the stabilization of the moduli at polynomial order 
requires a more careful treatment of the complete F-terms supplementing \eqref{eq:polextr} by additional constraints coming from metric-essential instantons. 

Let us outline an approach to read off these additional constraints by a more careful treatment of the F-terms. 
The exponentially suppressed contributions in the F-terms appear in the polynomial scalar potential through exponential eigenvalues of the K\"ahler metric, so it is convenient to expand $D_I W$ in an eigenbasis for the K\"ahler metric. The eigenvectors relevant for the supplementary constraints then have a vanishing eigenvalue under the polynomial part of the K\"ahler metric 
\begin{equation}
(\partial_i \bar{\partial}_{\bar{\jmath}}\log \cK_{\rm pol}) V^j = 0\, ,
\end{equation}
For boundaries with a linear $\cK_{\rm pol}$ such as in section \ref{ssec:II1} these eigenvectors $V^i$ will always be independent of the saxions, but for more general boundaries this need not be the case. Letting $a_{r_1 \cdots r_n}$ denote the essential instanton term corresponding to a given eigenvector $V^i$, then the relevant F-term constraint can be written as
\begin{equation}\label{eq:expextr}
V^i D_i W =V^i \partial_i W_{\rm inst} =  e^{2\pi i r_i t^i} \langle F_3 - \tau_* H_3, \, V^i (2\pi i  r_i + N_i) e^{t^i_* N_i} a_{r_1 \cdots r_n} \rangle = 0 \, ,
\end{equation}
where we used that $V^i \partial_i W_{\rm pol}=0$, and replaced the covariant derivative by a partial derivative because $V^i K^{\rm cs}_i \, W_{\rm inst}$ is subleading in the instanton expansion. Furthermore, note that we dropped correction terms subleading compared to the essential instanton $a_{r_1 \cdots r_n}$. The complete set of extremization conditions to stabilize the moduli at polynomial order of the scalar potential is then given by \eqref{eq:polextr} and as many equations of the form \eqref{eq:expextr} as there are metric-essential instantons. Since the exponential factor in \eqref{eq:expextr} is simply an overall factor, this yields a system of polynomial equations in the moduli.\footnote{This polynomial structure of the extremization conditions is natural from the perspective of the self-duality condition \eqref{eq:selfduality}. The dependence on the complex structure moduli then enters through the Hodge star operator, whose polynomial part is already non-degenerate as discussed below \eqref{eq:Wexpansion}. }

An interesting feature of the vacua constructed according to the above scheme is a natural hierarchy of mass scales. All moduli are stabilized by the polynomial part of the scalar potential, so the eigenvalues of the hessian $\partial_I \partial_J V$ will be polynomial as well. Recall that the inverse K\"ahler metric either has polynomial or exponentially large eigenvalues, and that the canonically normalized masses are then computed as the eigenvalues of $K^{IC} \partial_C \partial_J V$. This means that the moduli masses take polynomial or exponentially large values in the vevs of the saxions. Either way, this separates the mass scale of the complex structure moduli and the axio-dilaton from the exponentially small scale of $|W_0|^2$. This hierarchy makes our flux vacua particularly attractive from the perspective of the KKLT scenario \cite{Kachru:2003aw}, since it allows one to consistently integrate out the complex structure and axio-dilaton sector before dealing with the K\"ahler moduli. This point separates our construction from other findings \cite{Demirtas:2019sip,Blumenhagen:2020ire,Demirtas:2020ffz}, where the mass of the lightest complex structure modulus was of the same scale then $|W_0|^2$, requiring a more refined analysis. 

We end this general discussion by commenting on the application of our procedure near the different types of boundary points. Given a type $\rm I$ point our method does not work. Roughly speaking, this is due to the fact that there is too little information contained in the polynomial piece of the period vector and as a consequence there is not enough freedom in the fluxes to stabilize all moduli at the polynomial level of the scalar potential as a result. We elaborate on this point in  appendix  \ref{finite_distance_appendix}. Near type $\rm II$ points, we have a polynomial K\"ahler potential that is at most linear in the complex structure moduli, which makes it the simplest and most natural candidate for our procedure. In section \ref{ssec:II0} and \ref{ssec:II1}, we explicitly apply our method to the one and two moduli cases respectively. A type $\rm III$ point does not occur in complex structure moduli spaces of dimension less than three. As the models developed in \cite{Bastian:2021eom} only had a maximum of two moduli, we did not have an explicit realization of this scenario at our disposal. There is, however, no obvious reason for which the construction should not go through in that case. For a type $\rm IV$ that is not at LCS, the simplest scenario occurs for two moduli. In that case there is a technical obstruction to the success of our method, but this problem should be solvable given more than two moduli. These last two types deserve more concrete investigations in the future and we make some further comments about them in the conclusions.

\section{Explicit models}\label{sec:examples}
In this section we illustrate our method for finding flux vacua with exponentially small superpotentials by studying one- and two-moduli boundaries in complex structure moduli space. Our search uses asymptotic models for the periods near these boundaries constructed in \cite{Bastian:2021eom}, and we refer the reader to this work for more details about their construction and conventions. We emphasize that, since all possible one- and two-moduli boundaries have been classified \cite{Kerr2017,Grimm:2018cpv,Bastian:2021eom}, this construction yielded an exhaustive set of asymptotic periods. For completeness let us recall some aspects of this classification. For one modulus, we can have three main types I, II, IV at point-like boundaries. In the two-dimensional case the picture already 
becomes richer, since two one-dimensional boundaries can intersect in a point. At this intersection point again only the three types I, II, IV can occur. If this 
type is IV, then there are two classes of intersections: (1) the familiar large complex structure cases, (2) the so-called coni-LCS cases. 
Near both of these intersections exponentially small superpotentials have already been constructed by considering specific geometric examples in \cite{Demirtas:2019sip,Demirtas:2020ffz,Blumenhagen:2020ire}. We refer to these works for a detailed treatment. 
This leaves us with the intersections where one encounters the types I or II at the intersection point. The Type I boundaries are at finite distance and we will 
discuss in appendix \ref{finite_distance_appendix} that one cannot stabilize all moduli at the level of the polynomial scalar potential in these situations. This implies that we would need to include further 
non-essential instanton corrections to stabilize a remaining polynomially flat direction.  This leaves us with the class of $\mathrm{II}$ boundaries, and we investigate how to engineer exponentially small vacuum superpotentials near these boundaries in the following.

\subsection{One-modulus  asymptotic region near a type $\mathrm{II}$ point}\label{ssec:II0}
We start with the simplest case, namely the one-modulus asymptotic regions that are near type $\mathrm{II}$ boundaries. 
These have already been systematically studied in the math literature \cite{Tyurin:2003}, and such boundaries correspond to so-called Tyurin degenerations. The period vectors near these boundaries have been computed for an explicit example in the physics literature in \cite{Joshi:2019nzi}, and general models for these periods including essential instantons have been constructed in \cite{Bastian:2021eom}. Using the results from the latter, we find that the K\"ahler potential \eqref{eq:kahlerexpansion} takes the form
\begin{equation}\label{eq:kpII0}
\cK_{\rm pol}=4  y \, , \qquad \cK_{\rm inst} =  \frac{4  a^2(1+\pi y)}{\pi} e^{-4 \pi y} \, , 
\end{equation}
where $a \in \mathbb{R}_{\neq 0}$ is some model-dependent coefficient that controls the essential instanton term and we have dropped all the non-essential instanton corrections. The polynomial and exponential parts of the flux superpotential take the respective forms
\begin{equation}\label{eq:WII0}
\begin{aligned}
W_{\rm pol}&=-g_3-ig_4+(g_1+ig_2)t \, ,\\
W_{\rm inst}&=a \, e^{2\pi i t}(t-\frac{1}{\pi i})(g_1-ig_2)-a \, e^{2\pi i t}(g_3-ig_4)\, ,
\end{aligned} \,
\end{equation}
where the $g_i = f_i - \tau h_i$ are the components of the three-form fluxes with $\tau = c+is$ being the axio-dilaton. We now proceed to stabilizing the moduli at the polynomial level. The vanishing F-term conditions for the polynomial part of the flux superpotential can conveniently be written as
\begin{equation}\label{eq:DWII0}
\begin{aligned}
y D_t W_{\rm pol} - s D_\tau W_{\rm pol} &= (f_1+i f_2) y - (h_3+i h_4) s=0\, , \\
y D_t W_{\rm pol} + s D_\tau W_{\rm pol} &= -i(f_3+i f_4) -i  (h_1+i h_2) sy =0\, , 
\end{aligned}
\end{equation}
where we dropped exponentially suppressed terms and set the axions to zero, i.e.~$c=x=0$. We chose not to impose $W_{\rm pol}=0$ at first, leaving us with the full covariant derivatives at polynomial level in the above extremization conditions. The solution to these constraints is given by
\begin{equation}\label{eq:vevsII0}
s = \sqrt{-\frac{f_1 f_3}{h_1 h_3}}\, , \qquad y = \sqrt{-\frac{ f_3 h_3}{f_1 h_1}}\, , \qquad f_2 h_3 = f_1 h_4\, , \qquad f_4 h_1 = f_3 h_2\, , 
\end{equation}
where we require $f_1/h_3 >0$ and $f_3/h_1<0$ in order to get positive values for the saxions. Plugging this solution back into the superpotential \eqref{eq:WII0} we find that $W_{\rm pol} = 0$ holds when
\begin{equation}
h_2 =- f_1 \sqrt{-\frac{h_1 h_3}{f_1 f_3}}\, , \qquad h_4 = - f_3 \sqrt{-\frac{h_1 h_3}{f_1 f_3}}\, . \label{eq:II0Const2}
\end{equation}
Implementing the constraints \eqref{eq:vevsII0} and \eqref{eq:II0Const2}, we obtain flux vacua that have $W_{\rm pol}=0$. Thus, the scale of the superpotential is set by 
\begin{align}
|W|=|W_{\rm inst}| \sim a \, e^{- 2 \pi y} = a  \exp\Big(-2\pi \sqrt{-\frac{ f_3 h_3}{f_1 h_1}}\, \Big)\, .
\end{align}
We want to emphasize that the exponentially small scale of the superpotential is set by an essential instanton term that is required by consistency of the theory, i.e.~the coefficient $a$ cannot vanish for any geometric example. Furthermore, we can now relate the saxion masses and their vevs to the tadpole. The latter takes the form
\begin{equation}
Q_{\rm D3}=f_1 h_3 - f_3 h_1 \,.
\end{equation}
Computing the masses from the scalar potential as the eigenvalues of $K^{ac} \partial_c \partial_b V$ we find that they are given by the compact expression
\begin{align}
m_{t}^2=m_{\tau}^2=\frac{1}{\cV^2}(f_1 h_3 - f_3 h_1) =  \frac{Q_{\rm D3}}{\cV^2} \,,
\end{align}
where $m_t^2, m_\tau^2$ denote the canonically normalized moduli masses associated with the complex structure modulus and the axio-dilaton respectively. Let us also stress that this one-modulus case is rather non-generic as there are no metric-essential instantons required, which have to be present in the multi-moduli case as we will see below. Furthermore, the saxion vev $y$ is bounded from above by the tadpole as
\begin{align}
y \leq  2(f_1 h_3-f_3 h_1)= 2 Q_{\rm D3} \,,
\end{align}
which is saturated for $f_3=h_3$ and $f_1=h_1=1$. As the scale of $|W_0|$ is set by $e^{-2 \pi y}$, we cannot make it arbitrarily small by tuning the vev of the saxion $y$, i.e.
\begin{align}\label{eq:W0boundII0}
|W_0| \gtrsim e^{-2 \pi Q_{\rm D3}}\, .
\end{align}
 To be more concrete, we consider a specific set of fluxes that satisfy the above equations. Given $H_3 = (-1,-2,4,-2)$ and $F_3=(8,-4,8,16)$, we find that
\begin{equation}
s = 4\, , \qquad y=2\,, \qquad |W_0 | \sim a \, e^{-4\pi}\, ,\qquad m_t^2 = m_\tau^2 = Q_{\rm D3} =40\, ,
\end{equation}
where we dropped the volume factor $1/\cV^2$ in the masses for convenience.

\subsection{Two-moduli asymptotic region near a type II point}\label{ssec:II1}
Next we consider the more involved class of two-moduli  boundaries that intersect on a type II point. Such asymptotic regions are realized, for example, in the moduli space of the Calabi-Yau threefold in $\mathbb{P}^{1,1,2,2,6}_4[12]$ near a Seiberg-Witten point \cite{Kachru:1995fv, Curio_2001,Eguchi:2007iw,Lee:2019wij}. Using the asymptotic model from \cite{Bastian:2021eom}, we find that the K\"ahler potential \eqref{eq:kahlerexpansion} near these boundaries takes the form
\begin{equation}\label{eq:kpII1}
\begin{aligned}
\cK_{\rm pol} &= 4(y_1+n_2 y_2) \, ,  \\
\cK_{\rm inst} &= -2 a^2 e^{-4\pi y_2} \Big(  n_1 y_1+y_2 + \frac{1-n_1 n_2}{2\pi} \Big) \\
& \ \ \ \, -2 n_2 b^2 e^{-4\pi y_1} \Big( n_2(n_1y_1+y_2) -\frac{1-n_1 n_2}{2\pi} \Big) \\
& \ \ \ \, -4 ab e^{-2\pi y_1-2\pi y_2}\Big(  n_2 ( n_1 y_1+y_2) + \frac{(n_2-1)(1-n_2 n_2)}{4\pi} \Big) \cos(2\pi(x_1-x_2)) \, , \\
\end{aligned}
\end{equation}
where $a,b \in \mathbb{R}_{\neq 0}$ and $n_1,n_2 \in  \mathbb{Q}_{\geq 0}$ are model-dependent parameters that control the essential instanton terms. The polynomial and exponential parts of the flux superpotential take the respective forms
\begin{equation}\label{eq:WII1}
\begin{aligned}
W_{\rm pol} &=  ( g_1+ig_2) (t_1+n_{2}  t_2) +( g_1+i g_2)- (g_4+i g_5)\, ,\\
W_{\rm inst} &=   a \, e^{2\pi i t_2} \Big(ig_3\frac{1-n_1 n_2}{2 \pi } -g_3 n_{1} n_{2}  t_1 - g_3 n_{2} t_2-     g_6 n_{2}\Big) \\
  & \ \ \ +  b \, e^{2\pi i t_1} \Big(ig_3 \frac{n_{1} n_{2}-1}{2\pi}  -g_3 n_{1} t_1 -g_3 t_2 -
   g_6 \Big) +\mathcal{O}(e^{-4\pi y})\,  .
\end{aligned}
\end{equation}
 Let us already point out that the polynomial part of the K\"ahler potential and flux superpotential is very similar to those of the one-modulus $\mathrm{II}_0$ boundaries given in \eqref{eq:kpII0} and \eqref{eq:WII0}, for which we only need to replace $t$ by $t_1+n_2 t_2$ and relabel some of the fluxes. By computing the K\"ahler metric for \eqref{eq:kpII1} one sees that $\cK_{\rm pol}$ yields a degenerate polynomial metric, as elaborated upon below \eqref{eq:example_metric} . This degeneracy is cured by the exponentially suppressed term involving $a^2$ in $\cK_{\rm inst}$, which therefore is a metric-essential instanton. Note that the essential instanton term involving $b$ can also cure this degeneracy unless $n_2=0$. 

Now let us turn to the extremization conditions that have to be solved. The vanishing F-term conditions can conveniently be written as
\begin{equation}\label{eq:DWII1}
\begin{aligned}
(y_1 + n_2 y_2) D_{t_1} W - s D_\tau W &= (y_1+n_2 y_2)(f_1+if_2) -s(h_4+i h_5) = 0\, , \\
(y_1 + n_2 y_2) D_{t_1} W + s D_\tau W &=  -i f_4 +f_5 + s(y_1+n_2 y_2)(-ih_1+h_2)= 0\, , \\
\frac{n_2  D_{t_1} W - D_{t_2} W}{2\pi (a e^{-2\pi y_2}-n_2^2 b e^{-2\pi y_1} ) } &= i f_6 +s h_6 -(n_1 y_1 +y_2) (f_3 - i sh_3) = 0\, .
\end{aligned}
\end{equation}
The first two conditions take a similar form as \eqref{eq:DWII0} found near one-modulus $\mathrm{II}$ boundaries, where the saxion $y$ is now replaced by the linear combination $y_1+n_2 y_2$ and some fluxes are relabeled. Proceeding in the same fashion as the one-modulus setup, these two equations fix the saxions $y_1+n_2 y_2$ and $s$, and furthermore impose two relations on the fluxes that ensure their axionic partners have vanishing vevs
\begin{equation}\label{eq:solve1II1}
y_1+n_2 y_2 =\sqrt{-\frac{ f_4 h_4}{f_1 h_1}} \, , \qquad s =\sqrt{-\frac{f_1 f_4}{h_1 h_4}} \, , \qquad f_2 h_4 = f_1 h_5\, , \qquad f_5 h_1 = f_4 h_2\, ,
\end{equation}
where we require $f_1/h_4 >0$ and $f_4/h_1<0$ in order to get sensible values for the saxions. 

The relevance of the third equation in \eqref{eq:DWII1} is a bit more subtle. While this F-term is exponentially suppressed, it does contribute to the polynomial part of the scalar potential, in line with our discussion at the end of section \ref{ssec:asymptotics}. This can be seen by carefully inspecting the K\"ahler potential \eqref{eq:kpII1}, which yields a K\"ahler metric with an exponentially small eigenvalue for the eigenvector $(n_2, -1)$. This scaling then cancels against the scaling of the F-term when computing the scalar potential through \eqref{eq:potential}, hence contributing at polynomial order. Therefore we must require the third F-term given in \eqref{eq:DWII1} to vanish as well, which is solved by
\begin{equation}\label{eq:solve2II1}
n_1 y_1+y_2 =  \sqrt{-\frac{f_6 h_6}{f_3 h_3}}  \, , \qquad f_3 f_6 h_1 h_4 = f_1 f_4 h_3 h_6\, .
\end{equation}
where we must require $f_3 / h_6 >0$ and $f_6 / h_3 <0$ to have positive saxion vevs compatible with the positivity conditions given below \eqref{eq:solve1II1}.

Together \eqref{eq:solve1II1} and \eqref{eq:solve2II1} specify all vacua (with vanishing axion vevs) of the polynomial scalar potential arising from the flux superpotential \eqref{eq:WII1}. We are interested in constructing vacua with $W_{\rm pol}=0$, which amounts to additionally imposing
\begin{equation}
h_2 = -f_1 \sqrt{-\frac{h_1 h_4}{f_1 f_4}}\, , \qquad h_5 = -f_4 \sqrt{-\frac{h_1 h_4}{f_1 f_4}}\, .
\end{equation}
It is now instructive to compare the scale of the moduli masses and the vacuum superpotential again with the D3-brane tadpole. The tadpole contribution due to the fluxes can be split into two parts as
\begin{equation}
Q_{\rm D3}  = \underbrace{ f_1 h_4-  f_4 h_1}_{Q^{\rm pol}_{\rm D3} }   +\underbrace{\frac{1}{2}( f_3 h_6 -f_6 h_3)}_{Q^{\rm inst}_{\rm D3} }\, , 
\end{equation}
where we separated the fluxes that appear in the superpotential \eqref{eq:WII1} at polynomial order from the fluxes that enter at exponential order. From the positivity conditions on the fluxes given below \eqref{eq:solve1II1} and \eqref{eq:solve2II1} we see that all terms in the contributions $Q^{\rm pol}_{\rm D3}$ and $Q^{\rm inst}_{\rm D3}$ are positive. For the canonically normalized masses of the moduli we then compute the eigenvalues of $K^{ac}\partial_c \partial_b V$ and find that
\begin{equation}
m_\tau^2 = m_{t_1+n_2 t_2}^2 =  \frac{f_1 h_4-  f_4 h_1}{\cV^2} = \frac{Q^{\rm pol}_{\rm D3} }{\cV^2} \leq \frac{Q_{\rm D3}}{\cV^2}  \, ,
\end{equation}
where we dropped the overall volume factor $1/\cV^2$ for convenience. We do not write down the mass associated with the modulus $n_2 t_1 - t_2$ here, but note that it takes an exponentially large value in the fluxes, since this field corresponds to an exponential eigenvalue of the K\"ahler metric. 

For the scale of the vacuum superpotential we need to study the saxion vevs. By using \eqref{eq:solve1II1} and \eqref{eq:solve2II1} we find the following two bounds through the D3-brane tadpole
\begin{equation}
y_1+n_2 y_2 \leq \frac{1}{2}Q^{\rm pol}_{\rm D3}\, ,\qquad n_1 y_1+y_2 \leq Q^{\rm inst}_{\rm D3} 
\end{equation}
which are saturated for $f_1=h_1=1$, $f_4=-h_4$ and $f_3=h_3=1$, $f_6=-h_6$ respectively. The vacuum superpotential is then set by the smallest saxion vev as
\begin{equation}
|W_0| \sim e^{-2\pi \min(y_1,y_2)} \geq e^{-2\pi Q_{\rm D3}}\, .
\end{equation}
Given both this bound on the vacuum superpotential and \eqref{eq:W0boundII0} in the one-modulus setup, it is tempting to speculate that $|W_0|$ can be bounded from below by the D3-brane tadpole near any type II point, also in higher-dimensional moduli spaces. In light of the tadpole conjecture \cite{Bena:2020xrh,Bena:2021wyr} this hints at an interesting tradeoff, where making $|W_0|$ smaller by increasing $h^{2,1}$ runs into problems with achieving full moduli stabilization.

To make the above story more concrete, let us consider the Seiberg-Witten point in the moduli space of the Calabi-Yau threefold in $\mathbb{P}^{1,1,2,2,6}_4[12]$ as an example. Following \cite{Bastian:2021eom}, we find that this boundary corresponds to the monodromy data $n_1=0$, $n_2=1/4$. As flux quanta we pick
\begin{equation}
F_3 = (-4,-8,10,16,-8,-8)\, , \qquad  H_3 = (-2,1,1,-2,-4,5)\, .
\end{equation}
The scalar potential then has a minimum at
\begin{equation}
s = 4\, , \qquad y_1 = 1.5 \, , \qquad y_2 = 2 \, , \qquad |W_0| = 7.3\cdot 10^{-5}\, ,
\end{equation}
where we used that $b=-4 \Gamma(3/4)^4/(\sqrt{3} \pi^2)$, and ignored the coefficient $a$ in the expansion of the superpotential \eqref{eq:WII1} since $e^{-2\pi y_1} \gg e^{-2\pi y_2} $. The canonically normalized moduli masses are then computed to be
\begin{equation}
 m_{n_2 t_1-t_2}^2 = 1.3 \cdot 10^{10} \, , \qquad m_{t_1+n_2 t_2}^2  = m_{\tau}^2 = 40 = Q^{\rm pol}_{\rm D3} \, ,
\end{equation}
where we dropped the volume factor $1/\cV^2$ in the moduli masses for convenience.
Note in particular that there is one exponentially large mass, which corresponds to the eigenvector of the K\"ahler metric with an exponential eigenvalue. Furthermore, all moduli masses are orders of magnitude larger than the exponentially small $|W_0|^2$.

\section{Conclusions}\label{sec:conclusions}
In this paper, we 
described the systematic construction of Type IIB flux vacua with an exponentially small vacuum superpotential 
using recent insights from asymptotic Hodge theory. It is based on the fact that all asymptotic regions in complex structure moduli 
space can be classified \cite{Kerr2017} and that general models for asymptotic period vectors 
can be constructed in each of these regions \cite{Bastian:2021eom}. 
Crucially, this can be done without reference to a specific geometric realization and therefore gives a powerful tool 
to do model-building by directly manipulating the geometry of the moduli space. A crucial aspect of this construction is that the inclusion of  what we called 
essential instantons is important in most asymptotic regimes. These appear as exponential corrections to the polynomial part of the periods and are required  
 by the general statements of asymptotic Hodge theory. Their presence is also of physical significance, since they ensure, for example, that the kinetic terms
 of the complex structure moduli are non-degenerate and positive. In this work we have shown how essential  
 instanton corrections can be the crucial ingredient in obtaining vacua with an exponentially small 
 superpotentials and a hierarchy in the mass matrix of the moduli. 
 
 A first takeaway of our approach to constructing flux vacua is that there is a conceptual difference in whether 
 the exponentially small value of the superpotential is induced by essential or non-essential instantons. Non-essential instantons 
 are not required by consistency, but are well-known to arise in most known explicit geometric examples. They have 
 been central in the flux superpotential constructions of \cite{Demirtas:2019sip,Demirtas:2020ffz,Blumenhagen:2020ire} and lead to mass matrices with exponentially small eigenvalues. 
 In contrast, we showed that the exponential terms induced by essential instantons in the superpotential and K\"ahler potential 
 can combine to contributions to the scalar potential that enter at polynomial level. In fact, a special subset of these instantons, which 
 we termed \textit{metric-essential} since they ensure the non-degeneracy of the moduli metric, necessarily leads to such polynomial 
 corrections in the scalar potential. Remarkably, we were thus able to study moduli stabilization at polynomial level and an 
 exponentially small vacuum superpotential with the control provided by asymptotic Hodge theory. 

To concretely demonstrate how our method works we explicitly applied it to one- and two-dimensional complex structure moduli spaces by singling out a certain set of boundaries and associated asymptotic regions within in the classification of all occurring possibilities of \cite{Kerr2017}. The considered asymptotic regions were close to a Type II point in both 
the one- and two-dimensional moduli spaces, i.e.~we assumed that boundary of the highest co-dimension is of this specific type. 
We showed that in one- and two-dimensional moduli spaces this is a sufficient condition for implementing an exponentially small vacuum superpotential controlled by essential instantons and suspect that this conclusion also generalizes  to higher-dimensional moduli spaces. In the considered low-dimensional examples we have already observed some interesting features of these models. Firstly, we have seen that starting at dimension two the moduli metric always has exponentially small eigenvalues, while the scalar potential can still depend polynomially on the same field. This universal feature of Type II points in moduli space is only absent for the one-dimensional moduli spaces, which are known to be special also from other perspectives \cite{Grimm:2018ohb}. Secondly,  for both one- and two-dimensional moduli spaces, we have found that
one complex structure modulus and the axio-dilaton are stabilized at polynomial order and have a mass of the same order. 
Thirdly, we have shown that in the considered cases there is a simple bound on how small $|W_0|$ can be made. 
It is expected that such a bound exists due to the finiteness of 
flux vacua \cite{toappearMath,Grimm:2020cda}, but it is interesting that one can show that this bound is set by the exponential of the negative tadpole charge. Concluding this summary, let us note that we expect that most of these observations can also be established for higher-dimensional moduli spaces. 

Let us stress that our approach is of particular use in the realization of certain moduli stabilization scenarios, such as the KKLT scenario. The metric-essential instantons 
provide mass to the complex structure moduli that is naturally larger than the mass scale of the non-perturbatively stabilized K\"ahler structure moduli. 
This avoids the complications arising from dealing with a very light complex structure modulus. However, we note that it is a general feature of the outlined scenario in moduli spaces with $h^{2,1}>1$ that one finds at least one complex structure modulus that gains an exponentially large mass after canonical normalization. This 
is due to the fact that the moduli metric admits an exponentially small eigenvalue set by the metric-essential instantons. It needs to 
be ensured that the mass of this field is still sufficiently below the Kaluza-Klein scale to make our considerations self-consistent. 
One can then aim at implementing an exponentially small vacuum superpotential by including the outlined stabilization 
scheme as a building block in a more extensive setting. For example, one can stabilize a number of the 
complex structure moduli near a Type II boundary and use the essential instantons to control the superpotential. We 
expect that this can be done for explicit geometric Calabi-Yau examples or abstractly by using Hodge theory 
techniques. In either approach the described building block can be part of a general asymptotic moduli stabilization scheme put forward in \cite{Grimm:2019ixq,inprogress}.

There are numerous interesting questions for future research that might be answered using the general understanding of the asymptotic complex structure moduli space. Most immediate is to explore how our construction of flux vacua near Type II points extends to higher-dimensional moduli spaces and more involved boundary configurations. In particular, while we have presented arguments that our construction does not work for finite distance boundaries, there are more possibilities for infinite distance boundaries other than Type II and the large complex structure boundaries. For example, in higher-dimensional moduli spaces one can find novel Type $\mathrm{III}$ boundaries that cannot occur in complex structure moduli spaces of dimension less than three. Furthermore, there are also so-called coni-LCS boundaries, which roughly speaking arise when a finite distance boundary meets an infinite distance boundary. For those our method requires also at least three moduli in order to have enough freedom to stabilize moduli at the polynomial level. This fact ties nicely together with the recent construction of flux vacua near such boundaries in the three moduli setting that have been proposed in \cite{Demirtas:2020ffz,Blumenhagen:2020ire}. The most direct way to make progress on more general configurations is to extend the construction of asymptotic periods in \cite{Bastian:2021eom} and then use them to study flux vacua. Alternatively one can aim to apply the asymptotic Hodge theory techniques directly to the scalar potential as in \cite{Grimm:2019ixq,inprogress}, but this requires one to develop an efficient strategy to gain information about the superpotential. 

Let us close by stressing that the constructions of this work were carried out for Calabi-Yau threefolds with an understanding 
that an orientifold involution still needs to be implemented to obtain a minimally supersymmetric effective theory. A more elegant way to directly obtain an $\cN=1$ effective theory is to consider F-theory compactifications on elliptically fibered Calabi-Yau fourfolds. Remarkably, the used asymptotic Hodge theory methods are equally applicable to these higher-dimensional manifolds. In fact, we expect that these methods open the way to study interesting moduli stabilization schemes for F-theory as recently demonstrated in \cite{Grimm:2019ixq,Grimm:2020ouv} and \cite{Marchesano:2021gyv}. We believe that this novel way of studying F-theory flux vacua sets us on the track to understand the structure of the flux landscape, sidetracking the fact that only a sparse number of explicit period computations for Calabi-Yau fourfolds has been performed.  

\subsubsection*{Acknowledgments}
It is a pleasure to thank Jeroen Monnee, Jakob Moritz, Erik Plauschinn, and Lorenz Schlechter for very useful discussions and comments. 
This research is partly supported by the Dutch Research Council (NWO) via a Start-Up grant and a VICI grant.

\appendix

\section{Flux vacua near finite distance points} \label{finite_distance_appendix}

In this appendix we study flux vacua with exponentially small superpotentials near finite distance boundaries in complex structure moduli space, i.e.~type $\mathrm{I}$ points. We explain that the strategy laid out in section \ref{ssec:strategy} does not stabilize all moduli for such boundaries, and exponential corrections to the polynomial scalar potential $V_{\rm pol}$ are needed to lift flat directions.

Let us begin by inspecting the constraints put by the vanishing of the polynomial superpotential and the F-term of the axio-dilaton $\tau$ at the vacuum. These conditions together can be conveniently rewritten as
\begin{equation}
\langle F_3, \, \Pi_{\rm pol} \big|_* \rangle = 0 \, , \qquad \langle H_3,\,  \Pi_{\rm pol} \big|_* \rangle = 0 \, .
\end{equation}
Recall now that we can expand these polynomial periods as $ \Pi_{\rm pol} = e^{t^iN_i}a_0$, and subsequently use that $N_i a_0=0$ for type $\mathrm{I}$ points, giving $ \Pi_{\rm pol} = a_0$. It then follows that the polynomial part of the superpotential vanishes identically, i.e.~for any values of the axio-dilaton and the complex structure moduli we must have
\begin{equation}
W_{\rm pol} = \langle F_3 -\tau H_3, \, a_0 \rangle =  0\, ,
\end{equation}
since both $\langle F_3 , a_0 \rangle=0$ and $ \langle  H_3, a_0 \rangle  = 0$. Note that requiring $D_\tau W_{\rm pol}=0$ for these vacua thus only imposes a constraint on the fluxes $F_3$ and $H_3$, and in particular does not fix any of the moduli. Furthermore, corrections to this F-term through $D_\tau W_{\rm inst}$ only enter the scalar potential \eqref{eq:potential} at exponential order, since $K^{\tau \bar{\tau}}$ is polynomial in the fields. In contrast, metric-essential instantons cause the inverse K\"ahler metric $K^{i \bar{j}}$ to have $h^{2,1}$ exponentially large eigenvalues for finite distance boundaries, so the complex structure moduli F-terms $D_i W_{\rm inst}$ do contribute to the polynomial scalar potential. The polynomial scalar potential therefore reduces to
\begin{equation}
V_{\rm pol} = \frac{1}{\cV^2} e^K K^{i \bar{j}} D_i W_{\rm inst} D_{\bar{j}} \bar{W}_{\rm inst}\, ,
\end{equation}
where the indices $i,j$ run only over complex structure moduli. The crucial observation is now that the polynomial scalar potential is minimized by imposing $D_i W_{\rm inst}=0$. This yields only $h^{2,1}$ constraints on the complex structure moduli and axio-dilaton, leaving us with one free modulus. In principle one can lift this flat direction by including exponential corrections to the polynomial scalar potential $V_{\rm pol}$. In fact, to some extent we have control over these corrections, since non-metric essential instantons and $D_\tau W_{\rm inst}$ enter the scalar potential at exponential order. However, one might find that for instance non-essential instantons compete with these corrections to the scalar potential. Furthermore, even if the effective potential for this polynomially flat direction can be determined, the resulting extremization conditions will be of a much more complicated form than the algebraic equations \eqref{eq:polextr} and \eqref{eq:expextr} we encountered previously. For this reason we leave the search for flux vacua with exponentially small superpotentials near finite distance boundaries to future work.

\bibliographystyle{jhep}
\bibliography{references}

\providecommand{\href}[2]{#2}\begingroup\raggedright\begin{thebibliography}{10}

\bibitem{Grana:2005jc}
M.~Gra\~na, \emph{{Flux compactifications in string theory: A Comprehensive
  review}}, \href{http://dx.doi.org/10.1016/j.physrep.2005.10.008}{\emph{Phys.
  Rept.} {\bf 423} (2006) 91--158},
  [\href{https://arxiv.org/abs/hep-th/0509003}{{\tt hep-th/0509003}}].

\bibitem{Douglas:2006es}
M.~R. Douglas and S.~Kachru, \emph{{Flux compactification}},
  \href{http://dx.doi.org/10.1103/RevModPhys.79.733}{\emph{Rev. Mod. Phys.}
  {\bf 79} (2007) 733--796}, [\href{https://arxiv.org/abs/hep-th/0610102}{{\tt
  hep-th/0610102}}].

\bibitem{Giddings:2001yu}
S.~B. Giddings, S.~Kachru and J.~Polchinski, \emph{{Hierarchies from fluxes in
  string compactifications}},
  \href{http://dx.doi.org/10.1103/PhysRevD.66.106006}{\emph{Phys. Rev. D} {\bf
  66} (2002) 106006}, [\href{https://arxiv.org/abs/hep-th/0105097}{{\tt
  hep-th/0105097}}].

\bibitem{Kachru:2003aw}
S.~Kachru, R.~Kallosh, A.~D. Linde and S.~P. Trivedi, \emph{{De Sitter vacua in
  string theory}},
  \href{http://dx.doi.org/10.1103/PhysRevD.68.046005}{\emph{Phys. Rev. D} {\bf
  68} (2003) 046005}, [\href{https://arxiv.org/abs/hep-th/0301240}{{\tt
  hep-th/0301240}}].

\bibitem{Balasubramanian:2005zx}
V.~Balasubramanian, P.~Berglund, J.~P. Conlon and F.~Quevedo,
  \emph{{Systematics of moduli stabilisation in Calabi-Yau flux
  compactifications}},
  \href{http://dx.doi.org/10.1088/1126-6708/2005/03/007}{\emph{JHEP} {\bf 03}
  (2005) 007}, [\href{https://arxiv.org/abs/hep-th/0502058}{{\tt
  hep-th/0502058}}].

\bibitem{Grimm:2004uq}
T.~W. Grimm and J.~Louis, \emph{{The Effective action of N = 1 Calabi-Yau
  orientifolds}},
  \href{http://dx.doi.org/10.1016/j.nuclphysb.2004.08.005}{\emph{Nucl. Phys. B}
  {\bf 699} (2004) 387--426}, [\href{https://arxiv.org/abs/hep-th/0403067}{{\tt
  hep-th/0403067}}].

\bibitem{Bastian:2021eom}
B.~Bastian, T.~W. Grimm and D.~van~de Heisteeg, \emph{{Modelling General
  Asymptotic Calabi-Yau Periods}},
  \href{https://arxiv.org/abs/2105.02232}{{\tt 2105.02232}}.

\bibitem{Demirtas:2019sip}
M.~Demirtas, M.~Kim, L.~Mcallister and J.~Moritz, \emph{{Vacua with Small Flux
  Superpotential}},
  \href{http://dx.doi.org/10.1103/PhysRevLett.124.211603}{\emph{Phys. Rev.
  Lett.} {\bf 124} (2020) 211603},
  [\href{https://arxiv.org/abs/1912.10047}{{\tt 1912.10047}}].

\bibitem{Demirtas:2020ffz}
M.~Demirtas, M.~Kim, L.~Mcallister and J.~Moritz, \emph{{Conifold Vacua with
  Small Flux Superpotential}},  \href{https://arxiv.org/abs/2009.03312}{{\tt
  2009.03312}}.

\bibitem{Blumenhagen:2020ire}
R.~\'Alvarez-Garc\'\i{}a, R.~Blumenhagen, M.~Brinkmann and L.~Schlechter,
  \emph{{Small Flux Superpotentials for Type IIB Flux Vacua Close to a
  Conifold}},  \href{https://arxiv.org/abs/2009.03325}{{\tt 2009.03325}}.

\bibitem{Honma:2021klo}
Y.~Honma and H.~Otsuka, \emph{{Small flux superpotential in F-theory
  compactifications}},
  \href{http://dx.doi.org/10.1103/PhysRevD.103.126022}{\emph{Phys. Rev. D} {\bf
  103} (2021) 126022}, [\href{https://arxiv.org/abs/2103.03003}{{\tt
  2103.03003}}].

\bibitem{Demirtas:2021nlu}
M.~Demirtas, M.~Kim, L.~McAllister, J.~Moritz and A.~Rios-Tascon, \emph{{Small
  Cosmological Constants in String Theory}},
  \href{https://arxiv.org/abs/2107.09064}{{\tt 2107.09064}}.

\bibitem{Demirtas:2021ote}
M.~Demirtas, M.~Kim, L.~McAllister, J.~Moritz and A.~Rios-Tascon, \emph{{A
  Cosmological Constant That is Too Small}},
  \href{https://arxiv.org/abs/2107.09065}{{\tt 2107.09065}}.

\bibitem{Broeckel:2021uty}
I.~Broeckel, M.~Cicoli, A.~Maharana, K.~Singh and K.~Sinha, \emph{{On the
  Search for Low $W_0$}},  \href{https://arxiv.org/abs/2108.04266}{{\tt
  2108.04266}}.

\bibitem{Denef:2004ze}
F.~Denef and M.~R. Douglas, \emph{{Distributions of flux vacua}},
  \href{http://dx.doi.org/10.1088/1126-6708/2004/05/072}{\emph{JHEP} {\bf 05}
  (2004) 072}, [\href{https://arxiv.org/abs/hep-th/0404116}{{\tt
  hep-th/0404116}}].

\bibitem{Eguchi:2005eh}
T.~Eguchi and Y.~Tachikawa, \emph{{Distribution of flux vacua around singular
  points in Calabi-Yau moduli space}},
  \href{http://dx.doi.org/10.1088/1126-6708/2006/01/100}{\emph{JHEP} {\bf 01}
  (2006) 100}, [\href{https://arxiv.org/abs/hep-th/0510061}{{\tt
  hep-th/0510061}}].

\bibitem{schnell2014extended}
C.~Schnell, \emph{{The extended locus of Hodge classes}},
  \href{https://arxiv.org/abs/1401.7303}{{\tt 1401.7303}}.

\bibitem{Schmid}
W.~Schmid, \emph{{Variation of Hodge structure: the singularities of the period
  mapping}}, {\emph{Invent. Math. , 22:211--319, 1973} }.

\bibitem{CKS}
E.~Cattani, A.~Kaplan and W.~Schmid, \emph{{Degeneration of Hodge Structures}},
  {\emph{Annals of Mathematics} {\bf 123} (1986) 457--535}.

\bibitem{CKAsterisque}
E.~Cattani and A.~Kaplan, \emph{{Degenerating variations of Hodge structure}},
  in \emph{Th\'eorie de Hodge - Luminy, Juin 1987} (D.~Barlet, H.~Esnault,
  F.~Elzein, J.-L. Verdier and E.~Viehweg, eds.), no.~179-180 in Ast\'erisque,
  pp.~67--96.
\newblock Soci\'et\'e math\'ematique de France, 1989.

\bibitem{Kerr2017}
M.~Kerr, G.~Pearlstein and C.~Robles, \emph{{Polarized relations on horizontal
  SL(2)s}},  \href{https://arxiv.org/abs/1705.03117}{{\tt 1705.03117}}.

\bibitem{Grimm:2018cpv}
T.~W. Grimm, C.~Li and E.~Palti, \emph{{Infinite Distance Networks in Field
  Space and Charge Orbits}},
  \href{http://dx.doi.org/10.1007/JHEP03(2019)016}{\emph{JHEP} {\bf 03} (2019)
  016}, [\href{https://arxiv.org/abs/1811.02571}{{\tt 1811.02571}}].

\bibitem{Grimm:2019bey}
T.~W. Grimm, F.~Ruehle and D.~van~de Heisteeg, \emph{{Classifying Calabi-Yau
  threefolds using infinite distance limits}},
  \href{https://arxiv.org/abs/1910.02963}{{\tt 1910.02963}}.

\bibitem{Kachru:1995fv}
S.~Kachru, A.~Klemm, W.~Lerche, P.~Mayr and C.~Vafa, \emph{{Nonperturbative
  results on the point particle limit of N=2 heterotic string
  compactifications}},
  \href{http://dx.doi.org/10.1016/0550-3213(95)00574-9}{\emph{Nucl. Phys. B}
  {\bf 459} (1996) 537--558}, [\href{https://arxiv.org/abs/hep-th/9508155}{{\tt
  hep-th/9508155}}].

\bibitem{Eguchi:2007iw}
T.~Eguchi and Y.~Tachikawa, \emph{{Rigid limit in N=2 supergravity and
  weak-gravity conjecture}},
  \href{http://dx.doi.org/10.1088/1126-6708/2007/08/068}{\emph{JHEP} {\bf 08}
  (2007) 068}, [\href{https://arxiv.org/abs/0706.2114}{{\tt 0706.2114}}].

\bibitem{Gukov:1999ya}
S.~Gukov, C.~Vafa and E.~Witten, \emph{{CFT's from Calabi-Yau four folds}},
  \href{http://dx.doi.org/10.1016/S0550-3213(00)00373-4}{\emph{Nucl. Phys. B}
  {\bf 584} (2000) 69--108}, [\href{https://arxiv.org/abs/hep-th/9906070}{{\tt
  hep-th/9906070}}].

\bibitem{Grimm:2018ohb}
T.~W. Grimm, E.~Palti and I.~Valenzuela, \emph{{Infinite Distances in Field
  Space and Massless Towers of States}},
  \href{http://dx.doi.org/10.1007/JHEP08(2018)143}{\emph{JHEP} {\bf 08} (2018)
  143}, [\href{https://arxiv.org/abs/1802.08264}{{\tt 1802.08264}}].

\bibitem{Corvilain:2018lgw}
P.~Corvilain, T.~W. Grimm and I.~Valenzuela, \emph{{The Swampland Distance
  Conjecture for Kähler moduli}},
  \href{http://dx.doi.org/10.1007/JHEP08(2019)075}{\emph{JHEP} {\bf 08} (2019)
  075}, [\href{https://arxiv.org/abs/1812.07548}{{\tt 1812.07548}}].

\bibitem{Font:2019cxq}
A.~Font, A.~Herr\'aez and L.~E. Ib\'a\~nez, \emph{{The Swampland Distance
  Conjecture and Towers of Tensionless Branes}},
  \href{http://dx.doi.org/10.1007/JHEP08(2019)044}{\emph{JHEP} {\bf 08} (2019)
  044}, [\href{https://arxiv.org/abs/1904.05379}{{\tt 1904.05379}}].

\bibitem{Grimm:2019wtx}
T.~W. Grimm and D.~Van De~Heisteeg, \emph{{Infinite Distances and the Axion
  Weak Gravity Conjecture}},  \href{https://arxiv.org/abs/1905.00901}{{\tt
  1905.00901}}.

\bibitem{Grimm:2019ixq}
T.~W. Grimm, C.~Li and I.~Valenzuela, \emph{{Asymptotic Flux Compactifications
  and the Swampland}},
  \href{http://dx.doi.org/10.1007/JHEP06(2020)009}{\emph{JHEP} {\bf 06} (2020)
  009}, [\href{https://arxiv.org/abs/1910.09549}{{\tt 1910.09549}}].

\bibitem{Gendler:2020dfp}
N.~Gendler and I.~Valenzuela, \emph{{Merging the Weak Gravity and Distance
  Conjectures Using BPS Extremal Black Holes}},
  \href{https://arxiv.org/abs/2004.10768}{{\tt 2004.10768}}.

\bibitem{Lanza:2020qmt}
S.~Lanza, F.~Marchesano, L.~Martucci and I.~Valenzuela, \emph{{Swampland
  Conjectures for Strings and Membranes}},
  \href{https://arxiv.org/abs/2006.15154}{{\tt 2006.15154}}.

\bibitem{Grimm:2020cda}
T.~W. Grimm, \emph{{Moduli Space Holography and the Finiteness of Flux Vacua}},
   \href{https://arxiv.org/abs/2010.15838}{{\tt 2010.15838}}.

\bibitem{Bastian:2020egp}
B.~Bastian, T.~W. Grimm and D.~van~de Heisteeg, \emph{{Weak Gravity Bounds in
  Asymptotic String Compactifications}},
  \href{https://arxiv.org/abs/2011.08854}{{\tt 2011.08854}}.

\bibitem{Calderon-Infante:2020dhm}
J.~Calder\'on-Infante, A.~M. Uranga and I.~Valenzuela, \emph{{The Convex Hull
  Swampland Distance Conjecture and Bounds on Non-geodesics}},
  \href{https://arxiv.org/abs/2012.00034}{{\tt 2012.00034}}.

\bibitem{Grimm:2020ouv}
T.~W. Grimm and C.~Li, \emph{{Universal Axion Backreaction in Flux
  Compactifications}},  \href{https://arxiv.org/abs/2012.08272}{{\tt
  2012.08272}}.

\bibitem{Grimm:2021ikg}
T.~W. Grimm, J.~Monnee and D.~Van De~Heisteeg, \emph{{Bulk Reconstruction in
  Moduli Space Holography}},  \href{https://arxiv.org/abs/2103.12746}{{\tt
  2103.12746}}.

\bibitem{Lanza:2021qsu}
S.~Lanza, F.~Marchesano, L.~Martucci and I.~Valenzuela, \emph{{The EFT stringy
  viewpoint on large distances}},  \href{https://arxiv.org/abs/2104.05726}{{\tt
  2104.05726}}.

\bibitem{Castellano:2021yye}
A.~Castellano, A.~Font, A.~Herr\'aez and L.~E. Ib\'a\~nez, \emph{{A Gravitino
  Distance Conjecture}},  \href{https://arxiv.org/abs/2104.10181}{{\tt
  2104.10181}}.

\bibitem{Palti:2021ubp}
E.~Palti, \emph{{Stability of BPS States and Weak Coupling Limits}},
  \href{https://arxiv.org/abs/2107.01539}{{\tt 2107.01539}}.

\bibitem{Strominger_1995}
A.~Strominger, \emph{{Massless black holes and conifolds in string theory}},
  \href{http://dx.doi.org/10.1016/0550-3213(95)00287-3}{\emph{Nuclear Physics
  B} {\bf 451} (Sep, 1995) 96–108}.

\bibitem{doran2016mirror}
C.~F. Doran, A.~Harder and A.~Thompson, \emph{{Mirror symmetry, Tyurin
  degenerations and fibrations on Calabi-Yau manifolds}},  2016.

\bibitem{Tyurin:2003}
A.~N. Tyurin, \emph{{Fano versus Calabi--Yau}}, {\emph{The Fano Conference}
  (2004) 701--734}, [\href{https://arxiv.org/abs/math/0302101}{{\tt
  math/0302101}}].

\bibitem{Joshi:2019nzi}
A.~Joshi and A.~Klemm, \emph{{Swampland Distance Conjecture for One-Parameter
  Calabi-Yau Threefolds}},  \href{https://arxiv.org/abs/1903.00596}{{\tt
  1903.00596}}.

\bibitem{Curio_2001}
G.~Curio, A.~Klemm, D.~L\"ust and S.~Theisen, \emph{{On the vacuum structure of
  type II string compactifications on Calabi-Yau spaces with H-fluxes}},
  \href{http://dx.doi.org/10.1016/s0550-3213(01)00285-1}{\emph{Nuclear Physics
  B} {\bf 609} (Aug, 2001) 3–45}.

\bibitem{Lee:2019wij}
S.-J. Lee, W.~Lerche and T.~Weigand, \emph{{Emergent Strings from Infinite
  Distance Limits}},  \href{https://arxiv.org/abs/1910.01135}{{\tt
  1910.01135}}.

\bibitem{Bena:2020xrh}
I.~Bena, J.~Bl\r{a}b\"ack, M.~Gra\~na and S.~L\"ust, \emph{{The Tadpole
  Problem}},  \href{https://arxiv.org/abs/2010.10519}{{\tt 2010.10519}}.

\bibitem{Bena:2021wyr}
I.~Bena, J.~Bl\r{a}b\"ack, M.~Gra\~na and S.~L\"ust, \emph{{Algorithmically
  solving the Tadpole Problem}},  \href{https://arxiv.org/abs/2103.03250}{{\tt
  2103.03250}}.

\bibitem{toappearMath}
B.~Bakker, T.~W. Grimm, C.~Schnell and J.~Tsimerman, \emph{to appear}.

\bibitem{inprogress}
T.~W. Grimm, E.~Plauschinn and D.~van~de Heisteeg, \emph{to appear}.

\bibitem{Marchesano:2021gyv}
F.~Marchesano, D.~Prieto and M.~Wiesner, \emph{{F-theory flux vacua at large
  complex structure}},  \href{https://arxiv.org/abs/2105.09326}{{\tt
  2105.09326}}.

\end{thebibliography}\endgroup

\end{document}